\documentclass{aa} 
\pdfoutput=1
\usepackage{color}
\usepackage{amsmath}

\usepackage{graphicx}
\usepackage[varg]{txfonts}
\usepackage{natbib}
\usepackage{multirow} 	
\usepackage{multicol}

\usepackage{booktabs}
\makeatletter
\renewcommand*{\@fnsymbol}[1]{\ensuremath{\ifcase#1\or *\or \dagger\or \ddagger\or
    \mathsection\or \mathparagraph\or \|\or **\or \dagger\dagger
    \or \ddagger\ddagger \else\@ctrerr\fi}}
\makeatother

\DeclareUnicodeCharacter{2212}{-}
\begin{document}

\title{
Proper motion in lensed radio jets at redshift 3: a possible dual super-massive black hole system in the early Universe
  } 
\titlerunning{Proper motion in lensed radio jets}

\author{C.~Spingola\inst{1}\thanks{E-mail: spingola@astro.rug.nl} 
\and J.~P.~McKean\inst{1,2}
\and D.~Massari\inst{1,3,4}
\and L.~V.~E.~Koopmans\inst{1}}

\institute{Kapteyn Astronomical Institute, University of Groningen, Postbus 800, NL−9700 AV Groningen, the Netherlands
  \and ASTRON, Netherlands Institute for Radio Astronomy, Oude Hoogeveensedijk 4, 7991 PD Dwingeloo, the Netherlands 
  \and Dipartimento di Fisica e Astronomia, Universit\`{a} degli Studi di Bologna, Via Gobetti 93/2, I-40129 Bologna, Italy
 \and INAF - Osservatorio di Astrofisica e Scienza dello Spazio di Bologna, Via Gobetti 93/3, I-40129 Bologna, Italy}

\date{Received 7 March 2019 / Accepted 26 August 2019}

\abstract{In this paper, we exploit the gravitational lensing effect to detect proper motion in the highly magnified gravitationally lensed source MG~B2016+112. We find positional shifts up to 6 mas in the lensed images by comparing two Very Long Baseline Interferometric (VLBI) radio observations at 1.7 GHz that are separated by 14.359 years, and provide an astrometric accuracy of the order of tens of $\mu$as. From lens modelling, we exclude a shift in the lensing galaxy as the cause of the positional change of the lensed images, and we assign it to the background source. The source consists of four sub-components separated by $\sim 175$ pc, with proper motion of the order of tens $\mu$as yr$^{-1}$ for the two components at highest magnification ($\mu\sim350$) and of the order of a few mas yr$^{-1}$ for the two  components at lower magnification ($\mu\sim2$). We propose single AGN and dual AGN scenarios to explain the source plane. Although, the latter interpretation is supported by the archival multi-wavelength properties of the object. In this case, MG~B2016+112 would represent the highest redshift dual radio-loud AGN system discovered thus far, and would support the merger interpretation for such systems. Also, given the low probability ($\sim10^{-5}$) of detecting a dual AGN system that is also gravitationally lensed, if confirmed, this would suggest that such dual AGN systems must be more abundant in the early Universe than currently thought.} 

\keywords{Galaxies: active; Galaxies: jets; Gravitational lensing: strong; Instrumentation: high angular resolution; Instrumentation: interferometers; Radio continuum: galaxies }
	
	\maketitle 

\section{Introduction}
\label{sec:intro}
The formation of super-massive black holes (SMBHs) at the centres of galaxies is still an unclear process. According to the hierarchical structure formation scenario, SMBHs can be created as a result of a major merger of two galaxies, each with its own nuclear black hole \citep{Begelman1980,Volonteri2010}. Such systems can have an important effect on the build-up of the stellar halo through mechanical and radiative feedback when both black holes are undergoing an active phase. Also, the merger of such black holes may result in extreme gravitational wave events in the early Universe, which are the primary targets of the Laser Interferometer Space Antenna (LISA, e.g. \citealt{Enoki2004}). However, the lack of observed AGN pairs suggests that there must be a fast spiralling of the two black holes when they reach the final merging-stage at pc-scales \citep{Mayer2007}, and detecting such systems with 1 to 100~pc separation is extremely difficult, with only one pc-scale dual AGN system being confirmed to date \citep{Rodriguez2006}.  However, the low detection rate of active binary SMBHs seems to be in agreement with the theoretical expectation of dual AGN if only one of the two SMBHs accretes and emits radiation during the merger. Then, in order to become a double AGN, the system must undergo at least two other major mergers \citep{Volonteri2003}. Under these assumptions, numerical simulations based on the optical and X-ray emission from AGN show that the fraction of dual AGN increases from 0.1 per cent at $z=0$ to only a few per cent at $z=2$, \citep{Volonteri2016, Rosas-Guevara2018}. 

Observationally, the most common approach to identify such pairs of active SMBHs is to detect emission lines with an offset in velocity of a several hundred km\,s$^{-1}$. This velocity offset can be seen as a double peak in the lines that originate in the narrow line regions of the two AGN, if they are spatially unresolved (e.g. [O\,{\sc iii}] lines, \citealt{Liu2018}).  However, it is known that the double peak in the emission lines in AGN can be also due to a wide range of phenomena, like outflows, inflows and unresolved rotation of the gas surrounding the SMBH. Recently, thanks to integral field unit spectrographs, it has been revealed with high detail that the complexity of the emission line profile can be attributed to these phenomena in most of the cases \citep[e.g.][]{Roche2016,Davies2017}. Therefore, using the doubly peaked feature alone does not guarantee that the target is a dual AGN and a multi-wavelength approach is necessary to confirm the binary system. Complementary observations can be perfomed at X-rays, because the two SMBH should exhibit non-thermal X-ray emission and, therefore, are easy to recognize at these wavelengths \citep{Lena2018}. However, the limited angular resolution of X-ray instruments does not allow the identification of the closest pairs of AGN. If the two AGN are radio emitters, the high angular resolution of radio interferometers can spatially resolve the system. Therefore, radio interferometric observations provide one of the most direct methods to identify dual AGN \citep{Burke-Spolaor2018}.
 
In this context, gravitational lensing eases the confirmation of such close binary SMBH systems.  The magnifying effect of gravitational lensing can spatially resolve the two AGN, especially if they are located in the region at highest magnification, namely close to the caustics.  
However, the gravitational lensing effect is a rare phenomenon, as the probability of observing a multiply-imaged quasar is of the order $10^{-3}$ (e.g. Turner et al. 1984). Therefore, detecting a gravitationally lensed dual AGN source is expected to be extremely unlikely.
 
In this paper, we investigate the gravitational lensing system MG~B2016+112, whose peculiar properties have been puzzling since its discovery \citep[e.g][]{Garrett1994}. In particular, we compare two VLBI observations at 1.7 GHz separated by 14.359 years with the aim of better understanding the nature of the background radio source. We detect a significant positional shift in the lensed images between the two epochs, which can be interpreted as either a proper motion along the jets or an orbital motion of two radio-loud AGN in the source plane. In Section \ref{3.Sec:Target}, we introduce the radio properties of MG~B2016+112. A description of the VLBI observations and data reduction is provided in Section \ref{3.Sec:Observations}. We present the lens modelling and source reconstruction in Section \ref{3.Sec:Lens_modelling}, while the discussion and conclusions are presented in Sections \ref{3.Sec:Discussion} and \ref{3.Sec:Conclusions}, respectively.

Throughout this paper, we assume $H_0=67.8\; \mathrm{km\,s^{-1}~Mpc^{-1}}$, $\Omega_{\rm M}=0.31$, and $\Omega_{\Lambda}=0.69$ \citep{2018Planck}. The spectral index $\alpha$ is defined as $S_{\nu} \propto \nu^{\alpha}$, where $S_{\nu}$ is the flux density as a function of frequency $\nu$.

 \section{The target MG~B2016+112}
 \label{3.Sec:Target}
   
The gravitational lensing system MG~B2016+112 was discovered during the MIT-Green Bank survey (MG survey, \citealt{Lawrence1984, Bennett1986}). It consists of three images (A, B and C) of a background source at redshift $z=3.2773$, which is gravitationally lensed by an elliptical galaxy at redshift $z= 1.01$ and its satellite galaxy \citep{Lawrence1993, Yamada2001, Soucail2001, Koopmans2002}. 

From the first VLBI observations of MG~B2016+112 at 1.7 GHz, it was evident that images A and B are more compact, while image C is resolved into four sub-components connected by a faint extended emission \citep{Lawrence1984, Garrett1994,Garrett1996, Koopmans2002}. Later, high sensitivity and high angular-resolution observations with global VLBI at 1.7~GHz up to 8~GHz revealed that images A and B are resolved into 5 sub-components, where some sub-components have a flat spectral energy distribution, while others show a steep radio spectrum \citep{More2009}. Also, region C is resolved into multiple sub-components with both flat and steep radio spectra, where the two closest images, C12 and C22, show both compact and extended emission \citep{More2009}. Thanks to the high angular resolution of these VLBI observations, it was possible to measure that there is a significant asymmetry in the angular separation of the sub-components of the merging images in region C. The lensed images C11--C12 and C21--C22 should show a mirror inverted morphology and, therefore, should have the same angular separation and a similar flux density. Such an astrometric anomaly can be considered as an indication for a mass density perturbation, which in this case was attributed to the presence of a spectroscopically confirmed satellite galaxy (G1) that is south of region C \citep{KoopmansTreu2002, Chen2007, More2009}.

The lensing galaxy is an elliptical galaxy (called D) with a stellar velocity dispersion of $328\pm32$ km s$^{-1}$ \citep{KoopmansTreu2002}. Galaxy D is not active, as it does not display any emission at radio or X-ray wavelengths. This lensing galaxy lies in a massive galaxy cluster, which was detected for the first time at X-ray wavelengths \citep{Chartas2001, Toft2003}.

Several gravitational lens mass models have been proposed to reproduce the observations of MG~B2016+112 (e.g. \citealt{Narasimha1987, Nair1997}). Using the image positions given by early European VLBI Network (EVN) observations at 5 GHz, \citet{Koopmans2002} proposed a model where all of the lensed images belong to the same background source, which is assumed to be a core-jet AGN. In this model, the caustics pass through the AGN in a way that the core is doubly-imaged (region A and B) and part of the counter-jet (region C) is quadruply imaged. This model was revised and confirmed by \citet{More2009} using follow-up global VLBI observations at 1.7, 5 and 8.46~GHz. The multiple sub-components detected in the lensed images provided more constraints to the mass model. Moreover, \citet{More2009} included the faint satellite galaxy in the model, which accounts for most of the astrometric anomaly observed in region C. 

\section{Multi-epoch VLBI imaging}
\label{3.Sec:Observations}
In this Section, we present the new and archival VLBI observations used to investigate the proper motion of the lensed radio components.
\subsection{Observations}

MG~B2016+112 was observed with the ten 25-m antennas of the Very Long Baseline Array (VLBA) at a central frequency of 1.65 GHz (L-band) on 2016 July 2 (Epoch 2; ID: BS251, PI: Spingola). The experiment was carried out in phase-reference mode for a total observing time of 12 h. Scans on the phase-reference calibrator J2018+0831 of $\sim2$ min were alternated by observations of $\sim3.5$ min on the target (see Table \ref{Tab:Observations} for details). The fringe finder and bandpass calibrator for this experiment was 3C454.3. The data were recorded at 2 Gbit\,s$^{-1}$ and correlated at the VLBA correlator in Socorro to obtain two intermediate frequencies (IFs) of 128 MHz each, divided in 256 channels, at both circular RR and LL polarizations.

The archival global VLBI observations were taken on 2002 February 25 (Epoch 1; ID: GP0030, PI: Porcas), making the two epochs separated by $\Delta t =$ 14.359 years. The setup of the observation is summarized in Table \ref{Tab:Observations}. The fringe finder for these observations was B2029+121, which was also the phase-reference calibrator. The observations were correlated to obtain 4 IFs of 8 MHz bandwidth each, which were divided in 16 spectral channels. The observing antennas for this experiment were Effelsberg, Jodrell Bank, Medicina, Onsala, Torun, Robledo, Goldstone, all the antennas of the VLBA and the phased Very Large Array (VLA). Further details of these observations are reported by \citet{More2009}. For our analysis, we calibrate the VLBA-only subset of these observations, to match the {\it uv}-coverage between the two epochs.

\subsection{Data reduction}

We perform the calibration for both epochs following the standard VLBA procedures using the {\sc vlbautils} tasks built in the NRAO Astronomical Image Processing System ({\sc aips}; \citealt{Greisen2003}).  The amplitude calibration is based on the a priori knowledge of the system temperature and gain curve of each antenna. The initial calibration steps include corrections for instrumental offsets, Earth rotation, atmospheric opacity,  ionospheric dispersive delay and parallactic angle for the rotation of the antenna feed. Then, we correct for the fringe phases as a function of time and frequency (fringe fitting). Finally, we apply the bandpass calibration to correct for the response of the receiver as a function of frequency. All of these corrections are performed on the calibrators and then the solutions are interpolated to the target MG~B2016+112.

The imaging and self-calibration of both observations have been performed within the Common Astronomy Software Applications package ({\sc casa}; \citealt{CASA}). We apply phase-only self-calibration by starting with a solution interval as long as the scan length, which is iteratively decreased to 60~s. We do not use the first and last channels of the IFs. We use a Briggs weighting scheme during imaging (Robust = 0), which is a compromise between natural and uniform visibility weighting. The final restoring beam for the Epoch 2 observations is 11 mas $\times$ 5 mas at a position angle of $10$ degrees, while for the Epoch 1 (VLBA-only) observations is 11 mas $\times$ 4.5 mas at a position angle of $8$ degrees. Even if the difference in angular resolution between the two observations is small, it can lead to a possible incorrect identification of the  Gaussian centroids of the sub-components of the lensed images at the two epochs. In order to suppress the angular resolution effects in the estimate of the lensed image positions, we use the same weighting scheme and restoring Gaussian clean beam for imaging the target at the two epochs. This step allows us to recover the same angular scales and, therefore, identify the same sub-components in the lensed images. Nevertheless, the non-identical $uv$-coverage of the two observations may also lead to differences in the imaging. To minimize possible effects due to the different $uv$-coverage, we use only the VLBA antennas for imaging Epoch 1.

The off-source rms noise level is $\sim 70~\mu$Jy~beam$^{-1}$ for the first epoch and $\sim40~\mu$Jy~beam$^{-1}$ for the second epoch. This difference in sensitivity is to be expected, given the larger bandwidth of the new VLBA observations. The final self-calibrated VLBA images are shown in Fig.~\ref{Fig:images}. The observations from both epochs clearly resolve image A into two sub-components (A1 and  A2+A3), with an indication of two other sub-components (A4 and A5) that were detected in the global VLBI observations at 5 GHz by \citet{More2009}. The sub-components of image B are more distorted and blended together with respect to image A. Moreover, image C is resolved into four distinct sub-components at both epochs, and shows a faint extended emission in the tangential direction that connects images C12 and C22 in the observations taken during Epoch 2 (see Fig.~\ref{Fig:images}).

Finally, the total flux density of the system and the flux density of each sub-component are in agreement within the errors between the two epochs, indicating no significant flux density variability in this period at 1.7~GHz.

   \begin{table}
      \centering
      \caption{Summary of the VLBA observations at 1.7 GHz for MG~B2016+112 at Epoch 1 and Epoch 2.}
         \label{Tab:Observations}
  \begin{tabular}{lll}
  	\hline
       & GP0030 & BS251 \\
    \hline
       Date & 2002 February 25 & 2016 July 2  \\
       Instrument & global VLBI & VLBA \\
       On-source observing time & 8 h & 10 h \\
       IFs & 4 & 2 \\
       Total bandwidth & 32 MHz & 256 MHz\\
       Scans on target & 4.5 min & 3.5 min \\
       Scans on phase ref. & 2.5 min & 2 min \\
       Correlations & LL & RR, LL \\
	\hline        
    \end{tabular}
	\end{table}

\subsection{Measurement of the lensed image positions}

In order to measure the position of the lensed images, we fit the brightness distribution with two-dimensional Gaussian components using the task {\sc imfit} within {\sc casa}. Images A and B are fitted with two Gaussian components, while the four sub-components of image C could be fitted by a single Gaussian component each. Note that image A is resolved into at least three sub-components (see Fig.~\ref{Fig:images}). However, only two Gaussian components are clearly found when performing the fit (images A1 and A2+A3). As discussed in the previous section, the sub-components of image B are difficult to disentangle and clearly associate with the sub-components of image A. 
We use the Gaussian centroid as the position of the lensed images.  The uncertainty on the position is estimated in the standard way, and depends on the major and minor axes of the elliptical Gaussian and the signal-to-noise ratio, under the assumption that the component is unresolved. This is a major assumption, considering the significant blending of some sub-components at this frequency, which affects especially the lensed image B. Therefore,
the positional uncertainties estimated using this method should be considered as lower limits on the real ones. However, the two observations have almost the same {\it uv}-coverage, as we selected the same antennas (VLBA only) and the observing time is comparable. As a result, the morphology of the lensed images is found to be completely consistent between the two epochs (see Fig. \ref{Fig:images}).
For this reason, any deviation from a single 2D Gaussian function in the lensed images would be alike in the two epochs.\\

\subsection{Alignment of the two Epochs}

Self-calibration was fundamental for recovering both the compact and the extended structure of image C, which is at low surface brightness. However, this resulted in the precise absolute coordinate information being lost. Moreover, as the two observations have different phase-reference calibrators, this also resulted in a different absolute position of the lensed images at the two epochs. For these reasons, we need to align the two images to a common reference frame. This was picked as the frame defined by the self-calibrated image of Epoch 2. Positions measured in Epoch 1 are then transformed to the Epoch 2 reference frame by means of a linear transformation that is computed based on the flux density-weighted positions of A1 and B1 in the two epochs. Two sources are enough for this purpose, as VLBI observations provide a distortion-free reference frame, with the two images already having the same scale and rotation.\footnote{ We also searched for NVSS sources within the field-of-view that could be used for aligning the two images. However, the two closest NVSS sources are at $>6$ arcmin from the phase center and the distortion due to bandwith and time smearing is too strong to make them reliable astrometric references.}

 The position of the lensed images in the common reference frame and their uncertainties estimated with this method are listed in Table \ref{Tab:image-positions}.\\

\subsection{Positional offsets}

\begin{figure*}
   \centering
   \includegraphics[width= \textwidth]{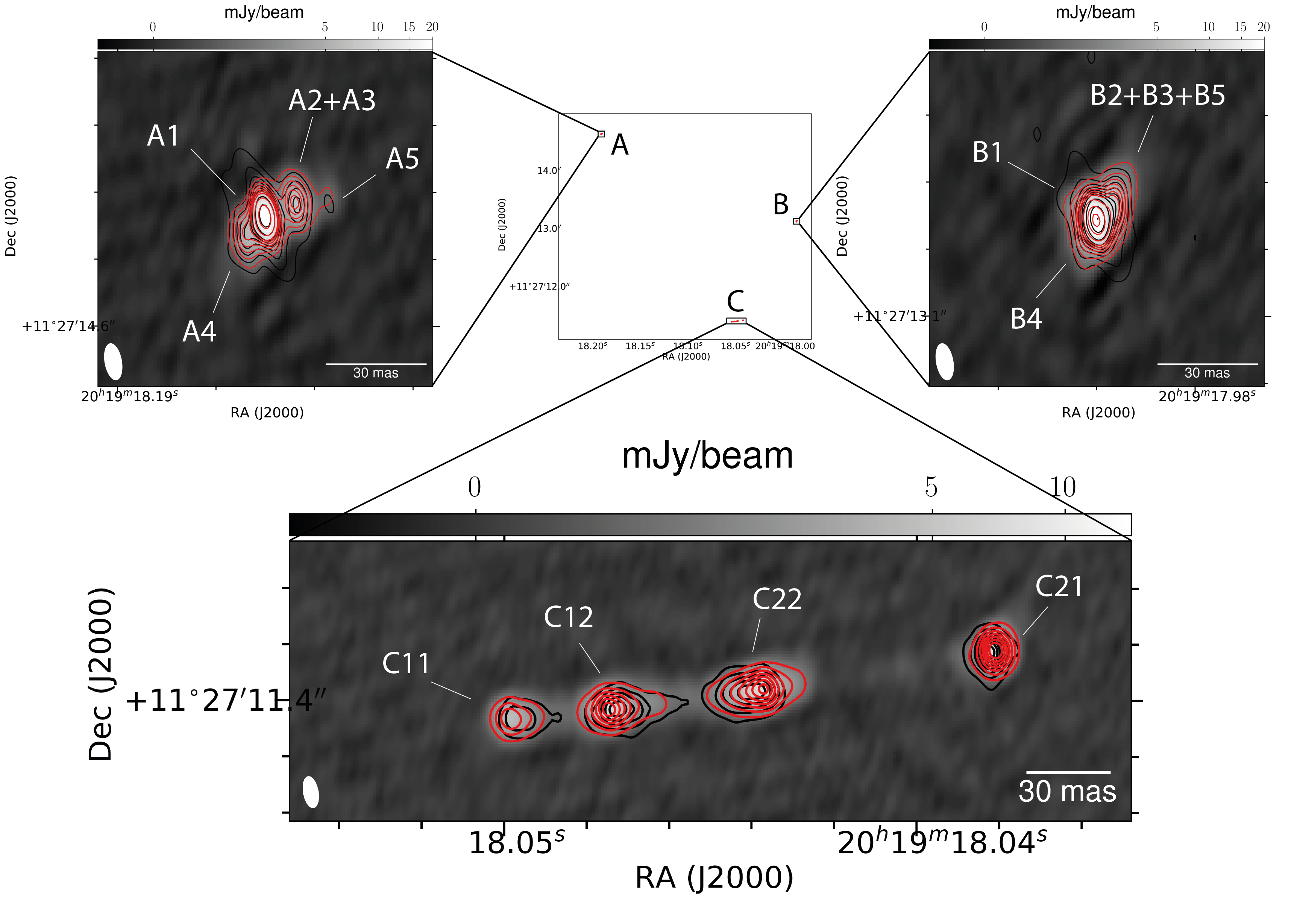}
   \caption{ Multi-epoch VLBA observations at 1.7 GHz of the gravitational lens MG~B2016+112. The central image shows the entire system as observed at 1.7 GHz with VLBI. The white contours are the observations taken on 2002 February 25 (Epoch 1), the greyscale map and the red contours are the new observations taken on 2016 July 2 (Epoch 2). The greyscale map is in units of mJy~beam$^{-1}$, as indicate by the colour bar in each image. On VLBI-scales, image A is resolved into four components (A1, A2+A3, A4 and A5 following the nomenclature of \citealt{More2009}), image B is resolved into two components (B1 and B2+B3+B5) with an indication for a possible third component (B4), while image C is resolved into four components (C11, C12, C22 and C21). The shift is more visible in region C, which is at higher magnification ($\mu \sim 270$ to $350$). Contours are at (0.05, 0.1, 0.15, 0.2, 0.25, 0.3, 0.35, 0.5, 0.75, 1) $\times$ the peak flux density of each individual image, which is $\sim22$ mJy~beam$^{-1}$ for Epoch 1 and $23$ mJy~beam$^{-1}$ for Epoch 2. The restoring beam is shown in the bottom left corner and is 11 mas $\times$ 5 mas at a position angle of $10$ degrees. All of the images are obtained using a Briggs weighting scheme, with Robust = 0.}
              \label{Fig:images}%
    \end{figure*}

We measure positional offsets in the range 0.4 to 5.7~mas in Right Ascension and 0.6 to  3 mas in Declination  between Epoch 1 and 2 for the various sub-components. These offsets are much larger than the astrometric uncertainties, which are between 8 to 30 $\mu$as for the group of sub-components associated with image C, and of the order of hundreds of $\mu$as for those making up images A and B (see Table \ref{Tab:image-positions}). The lensed images with the largest positional offsets are C11, C12, C22 and C21, which are also at the highest magnification region. The positional shift of this group of images is clearly visible in the image plane, as shown in Fig.~\ref{Fig:images}, and it is along the direction of highest magnification, which is a direct evidence for motion. Moreover, the direction of the shift is consistent with the symmetry expected by the gravitational lensing, that is, images C11--C12 and C21--C22 have moved in opposite directions along the highest magnification direction.

    \begin{table*}
   		 \centering
      		\caption[]{Position of the lensed images (Column 1) of MG~B2016+112 at Epoch 1 (Columns 2 and 3) and Epoch 2 (Columns 4 and 5) relative to the lensing galaxy,  which is at (0, 0). The position of the images is given by the centroid of the Gaussian fit performed using {\sc imfit}. The offsets in Right Ascension and Declination (Columns 6 and 7) are from the difference between Epoch 1 and Epoch 2.}
     	    \label{Tab:image-positions}
				  \begin{tabular}{ccccccc}
            \hline
            Image & $\alpha_1$ (arcsec) & $\delta_1$ (arcsec) & $\alpha_2$ (arcsec) & $\delta_2$ (arcsec)&  $\Delta \alpha$ (mas) & $\Delta \delta$ (mas) \\
            \hline
            \noalign{\vskip 0.15cm} 
   			 A1 & $-$1.74766$\pm$0.00014 & $+$ 1.77316$\pm$0.00038 	& $-$1.74769$\pm$0.00014& $+$1.77268$\pm$0.00030 			 		& $-$0.03$\pm$0.19 & $+$0.5$\pm$0.5 \\
             A2+A3 &  $-$1.73731$\pm$0.00025 & $+$1.77656$\pm$0.00146 	& $-$1.73769$\pm$0.00014& $+$1.77718$\pm$0.00030  															&$-$0.4$\pm$0.3 & $+$0.6$\pm$1.5 \\
 		 	 B1 &  $+$1.25914$\pm$0.00008 &  $+$0.27090$\pm$0.00019	& $+$1.25801$\pm$0.00003 &  $+$0.26999$\pm$0.00007															 & $-$1.1$\pm$0.1 & $-$0.9$\pm$0.2   \\
      		B2+B3+B5 &  $+$1.26013$\pm$0.00022  &  $+$0.27376$\pm$0.00103 & $+$1.26979$\pm$0.00009 &  $+$0.27075$\pm$0.00015														&  $+$9$\pm$3 & $-$3$\pm$1 \\
			C11 & $+$0.26659$\pm$0.00075  & $-$1.46016$\pm$0.00063 	 & $+$0.26245$\pm$0.00027 & $-$1.46123$\pm$0.00019															& $-$4.1$\pm$0.7 & $+$1.1$\pm$0.7\\
			C12 & $+$0.30288$\pm$0.00054	 & $-$1.45690$\pm$0.00029 	& $+$0.29720$\pm$0.00030 & $-$1.45748$\pm$0.00018 														 & $-$5.7$\pm$0.6 & $-$0.6$\pm$0.3\\
		    C22 & $+$0.34617$\pm$0.00046  & $-$1.45106$\pm$0.00028		 & $+$0.34885$\pm$0.00026 & $-$1.45082$\pm$0.00013														 & $+$2.7$\pm$0.5 & $+$0.2$\pm$0.3\\
            C21 & $+$0.43238$\pm$0.00020  & $-$1.43703$\pm$0.00025	 & $+$0.43306$\pm$0.00008 & $-$1.43697$\pm$0.00009 													 & $+$0.7$\pm$0.2 & $+$0.1$\pm$0.3\\
            \noalign{\vskip 0.15cm}    
            \hline
   	   \end{tabular}
	\end{table*}

\section{Lens modelling}
\label{3.Sec:Lens_modelling}

We model MG~B2016+112 using the software {\sc gravlens} and Monte Carlo realizations to estimate the errors on the mass model parameters and the source positions. To date, a change in the position of gravitationally lensed images over time has been only  detected in two doubly-imaged radio sources. A tentative detection of positional change in the lensed images has been observed in JVAS~B1030+074, where there is no clear correspondence among the source sub-components because the source lies in a region at low magnification \citep{Zhang2007}. While high frequency VLBI observations of  PKS~1830--211 revealed a change in the lensed images position that has been attributed to a motion in the source \citep{Jin2003}. Therefore, MG~B2016+112 represents one of the rare cases where proper motion has been clearly detected in a lensing system. Theoretically, the observed change in the image positions between Epoch 1 and 2 could be due to either a change in the lensing galaxy position (\citealt{Birkinshaw1989, Kochanek1996, Wucknitz2004b}) or a movement of one (or more) radio components in the source \citep{Jin2003,Biggs2005}. In this section, we explore both scenarios.

\subsection{The lensing galaxies have moved}

A possible explanation for the change in the image positions is that the lensing galaxy and/or the satellite galaxy have moved. For example, given that the lensing galaxy is in a cluster, with a velocity dispersion of $\sigma_v\simeq 771$~km\,s$^{-1}$ \citep{Soucail2001}, we would expect a positional change of just $\sim 1.5$ $\mu$as\footnote{As the proper motion $\mu$ is the distance traveled by the object divided by the time difference between the two epochs, the positional change (in arcsec) between the two epochs corresponds to $ v*\Delta t / (4.74*D)$; where $v = 771$ km\,s$^{-1}$, $\Delta t = 14.359$ years and $D = 1.576 \times 10^6$ pc.} in the 14 year period between Epoch 1 and 2. Although small, such a change in the position of the caustics could result in a significant change in the position of the lensed images, particularly for image C. 

In order to test how much the lensing galaxies could have moved from Epoch 1 to reproduce the image positions observed at Epoch 2, we proceed in the following way. By using the image positions at Epoch 1 and the lens mass model "scenario C" of \citet{More2009}, we keep all the lens mass model parameters fixed and we use the lensing galaxies positions produced as follows. We generate random positions for the lensing galaxies within a radius of 0.058 mas, which is the distance travelled by the galaxy if it moved at the speed of light for the temporal baseline $\Delta t$ between the two epochs, and is, therefore, a largely conservative assumption.
Since we do not have any information on the direction of motion that the two lensing galaxies (the main galaxy and its satellite) could have, the circles where we generate the positions for the lensing galaxies are uniformly filled. Then, we forward ray-trace the source components to the image plane for all the simulated positions of the lensing galaxies and determine if the predicted positions of the lensed images match the observations at Epoch 2.

We find that the model-predicted positions cannot fit simultaneously images A and B, and image C; either the model reproduces the doubly-imaged source or the quadruply-imaged source, with offsets between the observed and the model-predicted positions of the order of $30 \sigma$ on average; none of the simulated positions for the lensing galaxies can reproduce the position of the lensed images at Epoch 2. Therefore, we reject this scenario as a possible explanation for the positional shift observed in the lensed images between the two epochs.

 \begin{figure}
   \centering
   \includegraphics[width =0.5\textwidth]{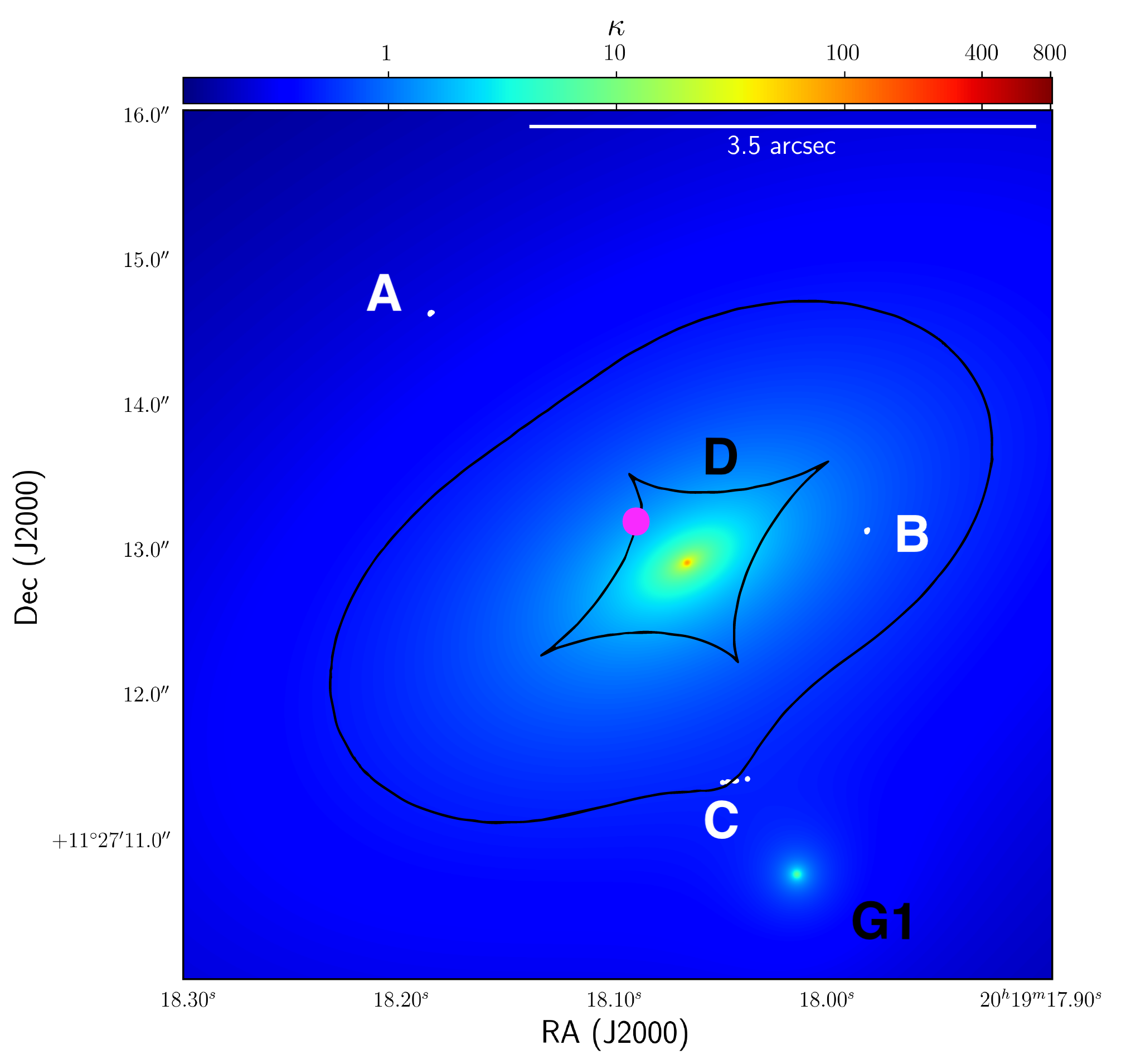}
   \caption{Convergence map for the lens mass model of MG~B2016+112. The field-of-view is 6 arcsec $\times$ 6 arcsec. The white contours are the 1.7 GHz observations at Epoch 2, while the black contours are the critical and caustic curves. The filled magenta circle indicates the location of the source components.
    The white labels indicate the groups of lensed images (A, B and C), and the black labels identify the two lensing galaxies (D and G1), also shown by the convergence peaks.}\label{Fig:lens-model}%
    \end{figure}
    
 \begin{figure*}
   \centering
   \includegraphics[width =\textwidth]{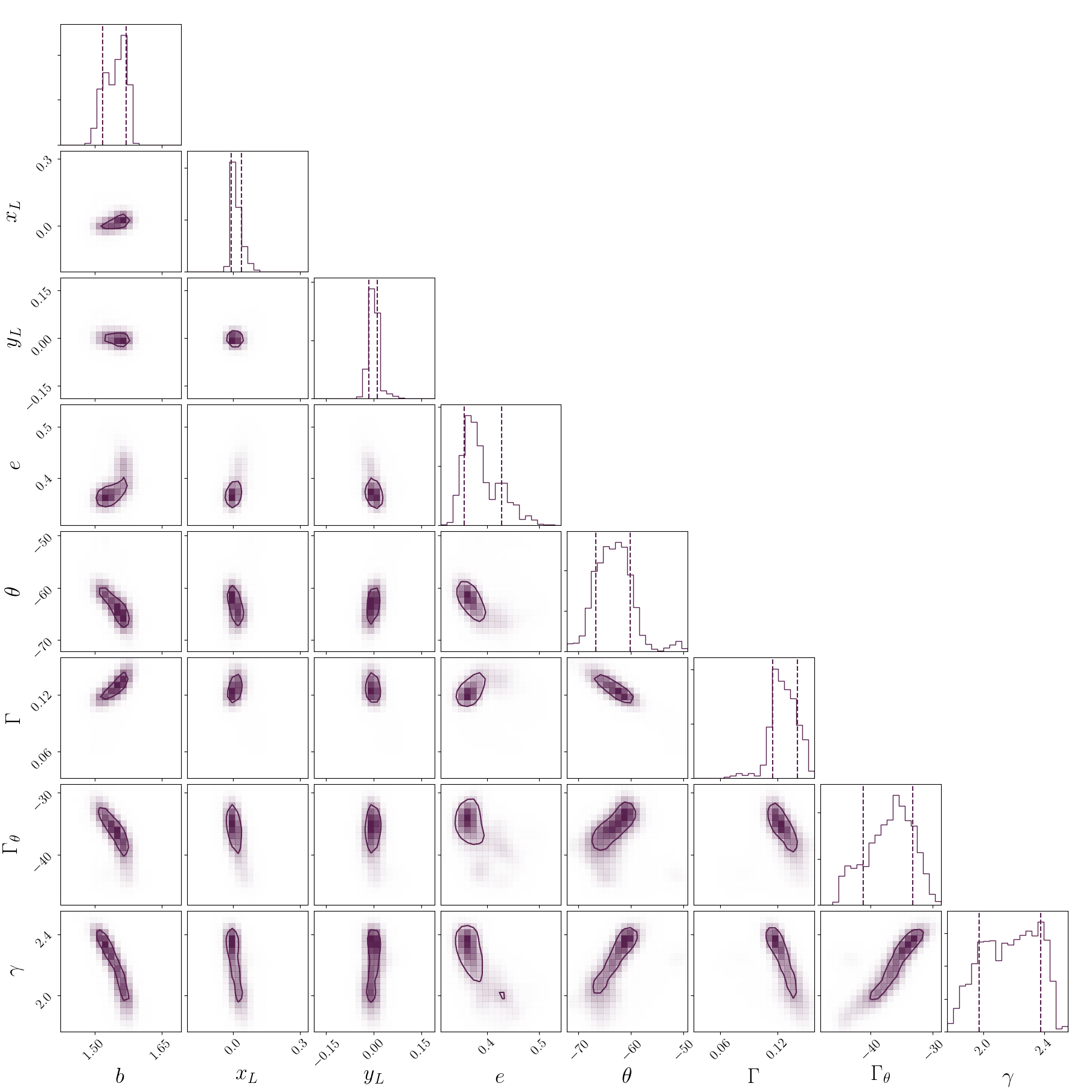}
   \caption{The diagonal histograms show the probability density functions (PDFs) for the lens model parameters of the main lensing galaxy (D), which were obtained by marginalizing over all the other model parameters, with two dashed vertical lines indicating the $1\sigma$ limits. The other panels show the 2-dimensional contours of the PDF for each pair of model parameters, where the contours indicate the $1\sigma$ region. The meaning of the parameters, their maximum-likelihood model value and their uncertainties are presented in Table~\ref{Tab:LensModel}.}\label{Fig:corner-plot}%
    \end{figure*}

\subsection{The source has moved}

The second scenario involves the source components in the source-plane moving with respect to the lensing galaxy position over the two epochs.
In order to reconstruct the position of the source components, we again assume the mass density distribution proposed by \citet{More2009} as a starting model. In particular, we adopt the model where images A1--B1 and A2--B2 are doubly imaged, while the pairs C11--C21 and C12--C22 are quadruply imaged (scenario C). We choose this model because the morphology of region C and the separation between the sub-components is typical of a pair of merging images in a four-image system (e.g. \citealt{Biggs2004}, \citealt{Hsueh2016}). The main lensing galaxy (D) is parameterized as an elliptical power law mass density distribution [$\rho(r) \propto r^{-\gamma}$]. The mass density distribution has 6 variables: the mass scale $b$; lens position $(x, y)$; ellipticity $e$ and its position angle, $\vartheta$, and power-law slope $\gamma$. We keep the satellite galaxy (G1) fixed, assuming the same mass model of \citet{More2009}. This model consists of a singular isothermal sphere (SIS) with mass strength $b = 0.145$ arcsec, at a position ($x_{G1}, y_{G1}$) = (0.840, $-$2.150) arcsec relative to the lensing galaxy D. We take into account the perturbation to the mass model due to the cluster of galaxies in the external shear term, which adds two other variables to the mass model, namely the shear strength $\Gamma$ and its position angle $\Gamma_{\vartheta}$.

We simultaneously use the position of the lensed images listed in Table \ref{Tab:image-positions} to constrain the mass density distribution. In this way, we are implicitly assuming that the same mass density distribution reproduces the lensed images at the two epochs.
This approach provides double the constraints to the lens model than using the two epochs separately, as each epoch provides an independent source distribution for the same lensing potential. Due to possible substructures within the lensing galaxy or along the line-of-sight, we do not use the relative flux-densities as additional constraints \citep{Metcalf2002, Mao2004, McKean2007, Rumbaugh2015, Hsueh2018, Despali2018}. Since the counter-images of components A4 and A5 are not resolved in image B, we do not use these components as constraints.

The best model parameters are presented in Table \ref{Tab:LensModel}, a schematic representation of the lens mass model is shown in Fig.~\ref{Fig:lens-model} and the probability density distribution for each parameter is shown in Fig.~\ref{Fig:corner-plot}. Our mass model did not deviate from the model proposed by \citet{More2009}, which already converged to a global minimum of the $\chi^2$ function.  In their model, \citet{More2009} fixed the ellipticity and position angle of the main deflector (D) to the values estimated from the surface brightness profile at near-infrared and optical wavelengths. Our model confirms that $e$ and $\vartheta$ are consistent with the parameters derived from the stellar emission within 1$\sigma$. Therefore, there is a good alignment between the stellar and the dark matter components within the Einstein radius, which is generally not observed for lens systems with a strong external shear ($\Gamma = 0.10\pm0.02$; \citealt{Spingola2018, Shajib2018}). We find the power-law slope to be consistent with an isothermal profile ($\gamma = 2.0 \pm 0.1$), which is consistent with the results obtained by \citet{Treu2002}, who combined gravitational lensing and stellar dynamics ($\gamma_{\rm TK2002} = 2.0 \pm 0.1$). 
 
Some of the model-predicted positions of the lensed images were found to differ from the observed positions by 2 to $10\sigma$, which was also noted by \citet{More2009}.  These positional residuals are not as critical as, for example, in the cases of CLASS B0128+437 and MG~J0751+2716 \citep{Biggs2004,Spingola2018}.
Therefore, the astrometric anomaly in MG~B2016+112 is not completely solved by the inclusion of G1, but could be due to an extra mass component that is currently not part of the model (e.g. see \citealt{Spingola2018} for discussion). \citet{More2009} also tested a model with three lensing galaxies, but found that this did not improve the model-predicted positions of the lensed images. Therefore, a more complex model for the mass density distribution is needed to fully explain the image positions of MG~B2016+112.

Our best model predicts the position and flux density of the counter-images of region C, at the position of region A and B, finding a flux density between 2 and 10 $\mu$Jy for the image pair C11--C21, and less than $1~\mu$Jy for the image pair C12--C22. These flux densities are lower by at least a factor of two when compared to the rms noise level of our imaging data. Therefore, the non-detection of these counter images in regions A and B is consistent with our best model, but future deeper observations that detect these counter images are needed to test the validity of the mass model.

\begin{figure}
   \centering
     \includegraphics[width =0.5 \textwidth]{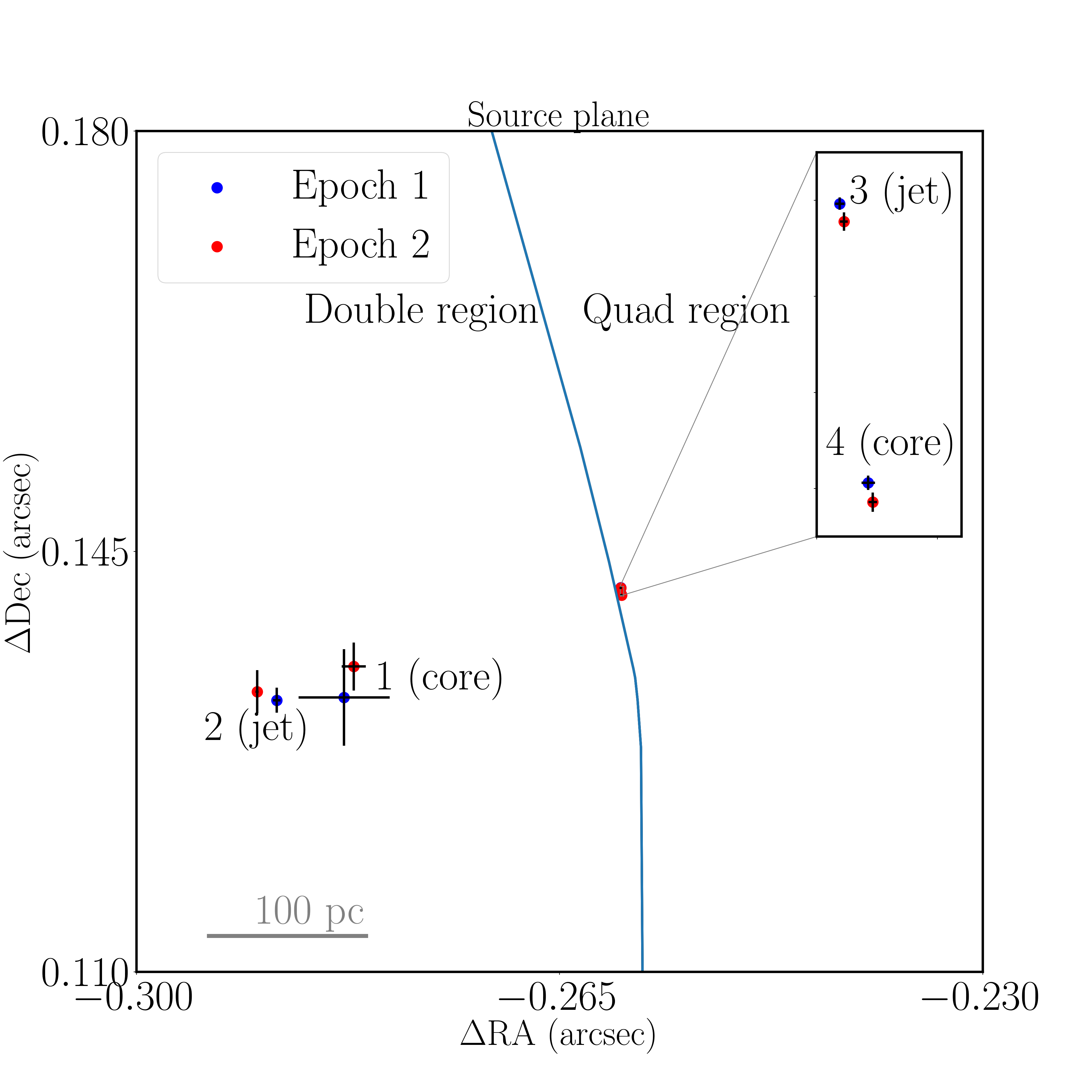}
   \caption{The source-plane model for the VLBA observations of MG~B2016+112. Epoch 1 (blue) and 2 (red) observations are aligned on the lensing galaxy position, which is at (0, 0). The solid blue line represents the caustics, which divides the source-plane into the double- and quadruple-image regions. The position of the sources is indicated by the filled circles. Source 1 corresponds to images A2--B2, source 2 corresponds to images A1--B1, source 3 corresponds to images C11--C21, and source 4 corresponds to images C12--C22. The labels "core" and "jet" are based on the radio spectral energy distribution of each image pair, as reported by \citealt{More2009}. The grey scale bar at the bottom left corner represents 100~pc at redshift $z=3.2773$. The error bars take into account the uncertainties in the lens model. In the case of sources 1 and 2 (related to the lensed images A and B), the error is dominated by our ability to de-convolve the different source components, while for sources 3 and 4 (related to the lensed region C), the errors are very small because the sub-components are clearly spatially resolved due to their extremely large magnification ($\mu \sim270$ to $350$).}\label{Fig:source-plane}%
    \end{figure}

In Fig.~\ref{Fig:source-plane}, we show the reconstructed source plane, given our best mass model, and we list the position of the source components in Table~\ref{Tab:physical-properties}. Source 1 corresponds to images A2--B2, source 2 corresponds to images A1--B1, source 3 corresponds to images C11--C21 and source 4 corresponds to images C12--C22. The uncertainty on each source position is estimated via Monte Carlo realizations using the following procedure. We simulate 1000 lensed images by randomly extracting them from a Gaussian distribution with a standard deviation that corresponds to the observed uncertainty and the expectation value of the observed position. We then keep our best lens mass model fixed and use these simulated lensed images to perform backward ray-tracing to obtain 1000 realizations for each source.  Finally, the uncertainty on each  component is given by the standard deviation of the 1000 source positions. In this way, the magnification ($\mu$) is taken into account in the estimate of the uncertainty, as opposed to just the uncertainty in the observed image position. As a result, regions with a higher magnification will have a lower positional uncertainty. Indeed, the source components associated with the highly magnified region C (sources 3 and 4; $\mu \sim 270$ and $\sim 350$, respectively) have a positional uncertainty of the order of 8 to 20 $\mu$as, while the astrometric uncertainty on the source components corresponding to images A and B (sources 1 and 2; $\mu \sim 2$ for both the sources) is between 1 and 4 mas (see Table~\ref{Tab:physical-properties}). The positional uncertainty is larger in Declination than in Right Ascension, which reflects the shape of the synthesized beam (see Fig.~\ref{Fig:images}). We highlight that source 1 has the largest positional uncertainty, not only because of the low $\mu$, but also because it is associated with the lensed image components A2--B2, which are the most difficult images to de-blend in the image plane, especially for image B (see Fig.~\ref{Fig:images}).

Our source reconstruction finds that components 3 and 4 have moved in the same direction (south) by $\sim40 \pm 25~\mu$as, while component 2 has shifted by $\sim2 \pm 1$~mas in the north-east direction. As shown in Fig.~\ref{Fig:source-plane}, the positional shift of source component 2 is statistically significant only in the Right Ascension direction. The positional shift in source components 3 and 4 is significant in both directions. 
We would like to emphasize that the lack of a significant change in flux density between the two epochs is consistent with our source reconstruction. This is particularly important for the source components at high magnification, as they are the most sensitive to any small change in the lensing configuration, due to the steep rise of the magnification when going close to the caustics. For example, if the source components associated with C11--C21 and C12--C22 moved towards (away from) the caustics, their flux density would have significantly increased (decreased) at the second epoch. Therefore, their measured flux densities are an indirect evidence that the motion of such components must have been parallel to the caustics. Finally, we note that any error on the lens model will not absorb the proper motion, but will modify the relative positions in the source-plane. 

   \begin{table}
      \centering
      \caption{The best recovered lens model parameters for the mass density distribution of the main deflector (D); $b$ is the mass strength in arcsec, $e$ is the ellipticity and $\vartheta$ is its position angle in degrees (east of north), $\Gamma$ is the external shear strength and $\Gamma_{\vartheta}$ is the shear position angle in degrees (east of north). The density slope is given by $\gamma$, where $\gamma = 2$ corresponds to an isothermal profile. We report the best set of parameters recovered (Best) via the minimization with {\sc gravlens} and the average values (Mean), with the relative 95 per cent confidence limit (CL), as assessed by the MCMC chains.}
         \label{Tab:LensModel}
  \begin{tabular}{lllll}
  	\hline
      Lens & Parameter & Best & Mean & $\sigma_{\rm mean}^{95 \%\, CL}$\\
    \hline
     \multirow{8}{*}{D}  & $b$ & 1.57 & 1.55 & $^{+0.02}_{- 0.03}$  \\
      & $\Delta$RA & 0.0 & 0.009 & $^{+0.029}_{- 0.015}$\\
       & $\Delta$Dec&  0.0 & $-$0.001 & $^{+0.013}_{- 0.014}$\\
       & $e$ & 0.43 & 0.38 & $^{+0.05}_{- 0.02}$ \\
       & $\vartheta$ & $-$59.1 & $-$63.3 & $^{+3.5}_{- 3.8}$ \\
	   & $\Gamma$ & 0.10 & 0.12 & $^{+0.02}_{- 0.01}$\\
       & $\Gamma_{\vartheta}$ & $-$41.5 & $-$36.5 & $^{+3.3}_{-4.7}$\\
       & $\gamma$ & 2.01 & 2.09 & $^{+0.09}_{-0.1}$ \\    

     \hline
        
    \end{tabular}
	\end{table}

   \begin{table*}
   \centering
      \caption[]{Properties of the source-plane components of MG~B2016+112, where the positions are measured relative to the lensing galaxy (at 0, 0). Given is the source component (Column 1), the Right Ascension and Declination at Epoch 1 (Columns 2 and 3), the Right Ascension and Declination at Epoch 2 (Columns 4 and 5), the offset in Right Ascension and Declination between Epoch 1 and Epoch 2 (Columns 6 and 7), and the proper motion in Right Ascension and Declination between Epoch 1 and Epoch 2 (Columns 8 and 9).}
         \label{Tab:physical-properties}
           \resizebox{\textwidth}{!}{%
  \begin{tabular}{ccccccccc}
  \hline
ID & $\alpha_1$ (arcsec) & $\delta_1$ (arcsec) & $\alpha_2$ (arcsec) &  $\delta_2$ (arcsec)  &  $\Delta\alpha$ (mas) & $\Delta\delta$ (mas) & $\mu_{\rm \alpha}$  (mas/year) &  $\mu_{\rm \delta}$ (mas/year)  \\
 \noalign{\vskip 0.15cm}
\cmidrule(lr){1-1}\cmidrule(lr){2-3}\cmidrule(lr){4-5}\cmidrule(lr){6-7} \cmidrule(lr){8-9} 
 \noalign{\vskip 0.15cm}
 1 (core) & $-$0.283$\pm$0.004 & $+$0.133$\pm$0.004 & $-$0.282$\pm$0.001 & $+$0.135$\pm$0.002 & $+$1$\pm$4 & $+$2$\pm$4 & $+$0.06$\pm$ 0.27 & $+$0.14$\pm$0.31   \\
  \noalign{\vskip 0.15cm}
\cmidrule(lr){1-1}\cmidrule(lr){2-3}\cmidrule(lr){4-5}\cmidrule(lr){6-7} \cmidrule(lr){8-9} 

 \noalign{\vskip 0.15cm}
 2 (jet) & $-$0.2884$\pm$0.0003 & $+$0.133$\pm$0.001 &  $-$0.290$\pm$0.001 &  $+$0.133$\pm$0.002 & $-$1.6$\pm$1.3& $+$0.5$\pm$2.1 & $-$0.11$\pm$0.08&  $+$0.04$\pm$0.13 \\
  \noalign{\vskip 0.15cm}
\cmidrule(lr){1-1}\cmidrule(lr){2-3}\cmidrule(lr){4-5}\cmidrule(lr){6-7} \cmidrule(lr){8-9} 

 \noalign{\vskip 0.15cm}
 3 (jet) & $-$0.25989$\pm$0.00001  & $+$0.14141$\pm$0.00001 & $-$0.259884$\pm$0.000009 & $+$0.14137$\pm$0.00002 & $+$0.006$\pm$0.015& $-$0.03$\pm$0.02 & $+$0.0005$\pm$0.0008 &  $-$0.002$\pm$0.001 \\
  \noalign{\vskip 0.15cm}
\cmidrule(lr){1-1}\cmidrule(lr){2-3}\cmidrule(lr){4-5}\cmidrule(lr){6-7} \cmidrule(lr){8-9} 
 \noalign{\vskip 0.15cm}
 4 (core) & $-$0.25995$\pm$0.00001 &  $+$0.14199$\pm$0.00002 & $-$0.2599434$\pm$0.000008 &  $+$0.14196$\pm$0.00001 & $+$0.008$\pm$0.013 & $-$0.04$\pm$0.02 & $+$0.0006$\pm$0.0008 &  $-$0.002$\pm$ 0.001  \\
  \noalign{\vskip 0.15cm}
 \hline
  \end{tabular}}
 \end{table*}

\section{Discussion}
\label{3.Sec:Discussion}

We have found evidence for proper motion in the lensed images of MG B2016+112 by analyzing two VLBI observations at 1.7 GHz that are separated by 14.359 years (see Fig.~\ref{Fig:images}). In Section \ref{3.Sec:Lens_modelling}, we ruled out the possibility that the proper motion is due to a shift in the lensing galaxy position, and we attribute it to a motion in the source. The source-plane reconstruction (see Fig.~\ref{Fig:source-plane}) shows that the de-lensed radio-loud object is quite complex, with two sets of two components moving in different directions. In this section, we investigate two possible interpretations for explaining the source morphology. First, we will explore the scenario where all of the sub-components belong to the same AGN. In this case, the motion is attributed to knots moving along the jets. Second, we will examine the possibility of having two separate radio-loud AGN in the source plane that are interacting with each other.

\subsection{Single AGN scenario}

Most jetted AGN show only one jet \citep{Padovani2017}. This is due to relativistic boosting, which enhances the radiation in the forward direction due to an approaching jet, and reduces the emission in the backward direction due to a receding jet \citep{Scheuer1979}. However, according to the unified AGN model, if the jet and counter-jet are seen under a large viewing angle, it is possible to detect both of them, as for example in FR I type radio galaxies \citep{UrryPadovani1995}. In these cases, it is expected that the counter-jet moves at sub-luminal velocities in an opposite direction with respect to the approaching jet, as seen for example in Centaurus A \citep{Jones1996}. 
MG~B2016+112 shows two optically thin components, which can be potentially associated with a jet and a counter-jet (sources 2 and 3, respectively),as one is moving at super-luminal velocity ($v_2$ = 2.9c$\pm$3.9c), while the other has a sub-luminal velocity ($v_3$ = 0.06c$\pm$0.04c) at an angular diameter distance of $D_A = 1576.2$ Mpc.

To test this scenario, we use the apparent motion of these two optically thin source components (sources 2 and 3) to estimate the theoretical ratio between the flux density of the possible jet and counter jet. This value can then be compared with the intrinsic flux density ratio between the source components associated with the jet and counter-jet, namely the flux densities corrected for the magnification. Since source component A1--B1 moves at a higher velocity than component C11--C21, we assume that source 2 consists of a knot in the approaching jet, while source 3 could be a knot in the receding counter-jet.

The ratio between the jet and counter jet flux densities can be written as 
\begin{equation}
R = \left(\frac{1 +\beta \cos(\theta)}{1 - \beta\cos(\theta)}\right)^{2+\alpha}
\end{equation}
where $\alpha$ is the spectral index, $\beta$ is the velocity expressed in units of $c$ and $\theta$ is the viewing angle \citep{Scheuer1979}. By assuming an intrinsically symmetric ejection of the two optically thin components, the factor $\beta \cos(\theta)$ can be expressed in terms of the proper motion of the jet ($\mu_{\rm j}$) and counter-jet ($\mu_{\rm cj}$),
\begin{equation}\label{Eq:beta_costheta}
\beta \cos(\theta) = \frac{\mu_{\rm j} - \mu_{\rm cj}}{\mu_{\rm j} + \mu_{\rm cj}}
\end{equation}
\citep{Fender1999}.

From Eq.~(\ref{Eq:beta_costheta}), assuming $\beta = 1$ and $\alpha = 0.8$\footnote{This is the average spectral index between 1.7 and 5 GHz for the optically thin components \citep{More2009}.}, we find a maximum viewing angle of $\theta_{\rm max}\simeq 17$ degrees, and a theoretical flux density ratio of $R \simeq 37\,500$. However, the observed flux density ratio, when corrected for the lensing magnification, between the jet (A1--B1) and the counter-jet (C11--C21) is $\sim 270$, which is two orders of magnitude less than the predicted ratio. However, this criterion is based on the strong assumption of a symmetric ejection of the knots in the jet and counter-jet. Moreover, given the large light travel time between jet and counter-jet, and the likely not simultaneous changes in the two radio ouflows, our $R$ value should be taken only as an indication that the single AGN scenario may not be completely compatible with the observations, rather than a conclusive statement. Also, the projected direction of the motion indicates that the two flat-spectrum components (i.e. the cores; sources 1 and 4) are moving perpendicularly to each other (see Fig.~\ref{Fig:source-plane}), even though the positional uncertainty is large for source component 1. This motion would imply an exotic jetted-AGN, or a possible reverse shock in the emission at pc-scales, as observed for example in the powerful radio jets of M87 and 3C345 \citep{Unwin1992}. 

\subsection{Dual AGN scenario}

The multi-wavelength properties of MG~B2016+112, when taken together, are also consistent with a possible dual AGN (DAGN) interpretation (defined as a pair of AGN separated by less than 10 kpc, while binary AGN consists of a pair of SMBH that are separated by less than $<100$ pc, \citealt{Burke-Spolaor2014}). DAGN show specific morphological and spectral features, such as multiple flat-spectrum radio-cores and misaligned/disturbed kpc-scale jets with a S- or X- shaped morphology \citep[e.g.][]{Deane2014, Burke-Spolaor2018}; jet-dominated radio emission \citep{Frey2012, An2018}; double peaked optical spectral emission lines separated by a few hundred km\,s$^{-1}$ (e.g. H$\beta$, [O{\sc iii}], \citealt{Comerford2009, Liu2018}); and multiple X-ray point source components \citep{Koss2012}.

\subsubsection{Evidence in favour of the DAGN scenario from previous studies}

 \citet{Yamada2001} showed that the narrow-line spectra of images B and C have different properties. These spectra, obtained with the Canada-France-Hawaii Telescope (CFHT), show typical emission lines from active galaxies (e.g. C\,{\sc iii}, C\,{\sc iv}, He\,{\sc ii}, N\,{\sc v}). \citeauthor{Yamada2001} found that they could not fit a single photo-ionization model that could explain simultaneously the line ratios with He\,{\sc ii} and those with N\,{\sc v} for both images B and C. This led to the interpretation that the excitation between these different parts of the background source is also different, concluding that MG~B2016+112 is likely a partially dust-obscured low-luminosity narrow-line AGN.

These differences in the optical spectra of images B and C could be due to two separate narrow-line regions, one associated with an un-obscured AGN (images A and B, which show a quasar morphology at optical wavelengths) and the other associated with a dust-obscured AGN (image C, which has an extended optical morphology). If so, the same emission lines should show a velocity offset. However, the low spectral-resolution of the CFHT observations does not provide an accurate enough measurement of the relative velocities of the narrow lines in images B and C, and the line properties of image A are still unknown. Alternatively, as region C is close to the caustics, the difference in the emission line flux-ratios could be due to a large differential magnification across a complex narrow line region. Therefore, even though the different line flux-ratios could be interpreted as evidence for a DAGN, further observations to measure the relative line velocities and their positions relative to the lensing caustics are needed.

Observations with \textsl{Chandra} showed that all three lensed images are X-ray sources \citep{Hattori1997, Chartas2001}. When correcting for the distortion due to gravitational lensing, the source corresponding to images A and B has a 2--10 keV luminosity of $3\times10^{43}$--$1.4\times10^{44}$ erg s$^{-1}$, but the authors do not investigate the intrinsic X-ray properties of the source related to image C. The images are quite faint in X-rays, with only 6, 5 and 12 photon counts for images A, B and C, respectively \citep{Chartas2001}.

The detection of multiple X-ray components associated with images A and B, and image C \citep{Chartas2001}, which also may have different intrinsic luminosities, could be explained by the presence of two possible distinct accretion disks within MG~B2016+112. Nevertheless, \citet{Chartas2001} explain these differences in the X-ray properties as images A and B being associated with the AGN, and the emission from region C being related to inverse Compton emission associated with the radio jets. However, from the current data, it is not clear which interpretation is correct for the X-ray emission from MG B2016+112.

Based on the radio spectral energy distribution \citep{More2009}, there is evidence of two flat-spectrum components and two steep-spectrum components. Classically, the flat-spectrum component is considered the core (i.e. the emission at the base of the jet, closest to the black hole), while the steep-spectrum component consists of the jet(s) of the AGN. Therefore, there are two possible cores and two possible jets in the source plane of MG~B2016+112. These two candidate core-jet AGN are intrinsically faint, with flux densities of the individual sub-components between $0.01$ and $10$ mJy. These properties at radio wavelengths can be taken as evidence in favour of the DAGN scenario.

\subsubsection{Evidence from proper motion}

The measurement of proper motion, and the direction of this proper motion in the source plane for the two different parts of the background source can also be taken as evidence for the DAGN interpretation. The source-plane consists of four components; sources 1 and 4 are the two flat-spectrum components ($\alpha \sim 0.2$ between 1.7 and 5 GHz), while sources 2 and 3 have a steeper spectral index ($\alpha \sim 0.8$ between 1.7 and 5 GHz; \citealt{More2009}). Given their relative projected distance in our reconstructed source-plane (see Fig.~\ref{Fig:source-plane}), they seem to form two separate core-jet structures. Therefore, we associate sources 1 and 4 with candidate radio cores, while sources 2 and 3 are identified as candidate jet components, as discussed briefly in the previous section. The separation between the two core-jet AGN is about 175~pc in projection, which is a strong indication that the two objects should be gravitationally bound, potentially forming a DAGN. The relative position of the optically thin components seems to indicate a misalignment between the radio jets. The presence of two flat-spectrum components and multiple misaligned jets is generally a criterion used for identifying DAGN at radio and X-ray wavelengths \citep{Owen1985, Lal2007}, making this morphology consistent with the DAGN scenario. Clearly, more precise positional measurements of the source components 1 and 2 are needed to confirm the differences between the jet alignment.

The most unusual feature of the core component associated with images C12--C22 is the proper motion (source 4; see Fig.~\ref{Fig:source-plane}).  Generally, the core is stationary in single AGN galaxies \citep[e.g][]{Marscher2009}. Therefore, a movement of the radio core, as may be seen here, would imply a shift of the entire AGN system. Moreover, the two optically thin components (especially source 3) are moving in a similar direction as their associated core components. This could be due to the two AGN dragging their jets while they move. Sources 3 and 4 are found to be moving with an apparent sub-luminal velocity in the southern direction, with $v_3 = 18900 \pm 15000$ km s$^{-1}$ and $v_4 = 20100 \pm 13000$ km s$^{-1}$, for the candidate jet and core, respectively. Source 1 does not show significant motion within the uncertainties (see Table \ref{Tab:physical-properties} and Fig.~\ref{Fig:source-plane}), while source 2 is moving with an apparent super-luminal velocity of $v_2 = 2.9c \pm 3.9c$. This velocity indicates the presence of Doppler boosting, and hence, requires the jet to be oriented at a small viewing angle. Therefore, the motion of this component might be a combination of the proper motion of the entire AGN, and the motion of the optically thin outflow with respect to the main core. Even if the velocities of the two AGN are higher than those typical of galaxies in clusters ($<1000$ km$\,$s$^{-1}$), our findings are still consistent with a scenario where the two AGN candidates are at a later stage of the merging and, therefore, accelerating. This is even more plausible given their small angular separation.

We find that both candidates AGN show low intrinsic flux densities, but they have different radio properties \citep{More2009}. The flux density of the possible AGN comprising sources 1 and 2 is dominated by the emission from the jet, whereas the candidate AGN composed of sources 3 and 4 is core-dominated. This kind of difference between the two radio-loud SMBH can be attributed to the different orientations of the two interacting AGN, which may be further evidence in favour of the DAGN scenario.

\subsection{Implications of a dual AGN scenario}
To date, two main avenues have been proposed for the formation of SMBHs in galaxies: the accretion of gas from a directly collapsed star \citep{Begelman2002} or the merging of multiple black hole seeds \citep{Volonteri2003}. The presence of a DAGN in MG~B2016+112 would be in favour of the merger-driven formation scenario, as DAGN represent an intermediate evolutionary stage of such a process. According to the hierarchical formation scenario, this is expected to be observed at high redshift \citep{Volonteri2016}. The timescales on which multiple SMBHs can coalesce are not known, but it is expected to be short given the low detection rate of such systems. Therefore, observations of small separation DAGN are needed in order to probe the final stages of the merging process. 

As the lensing probability of radio-loud AGN is $\sim10^{-3}$ and the expected fraction of AGN with a dual SMBH system is $\sim10^{-2}$ at high redshift, the combined probability of detecting a gravitationally lensed DAGN system is of order $10^{-5}$. Therefore, the detection of a gravitationally lensed DAGN here would imply that the overall probability is higher than expected, meaning that DAGN are more common in the early Universe than first thought. As the MG survey detected $\sim$6000 sources at signal-to-noise ratio above 5 \citep{Bennett1986} and found six strong gravitational lensing systems, this gives approximately a strong lensing probility of 10$^{-3}$, which is a typical strong lensing rate \citep[e.g.][]{Spingola2019}. If MG~B2016+112 is a genuine DAGN, then we can roughly assume that 1 every 6 strongly lensed sources is a DAGN, resulting in a probability of detecting a gravitationally lensed DAGN of 0.16 per cent. Therefore, the probability of finding a gravitationally lensed DAGN system in the MG survey is $\sim$2$\times 10^{-4}$, which is an order of magnitude higher than the expectations from the current simulations.

To some extent, this is expected given the increased merger rate  (expected at high redshift), but to catch one in the act of forming would also mean that the in-spiral time needs to be slower than what is currently predicted \citep[e.g.][]{Rafikov2016}. As this conclusion is currently based on the statistics of this single possible detection, further studies of other lensed radio sources at high angular resolution with VLBI are needed to determine if there are other cases with proper motion in the radio jets, and whether such motion is also consistent with a DAGN system.

\section{Conclusions and future work}
\label{3.Sec:Conclusions}

In this paper, we have presented a clear detection of proper motion from a gravitationally lensed radio source at high redshift. From analyzing two VLBI datasets separated by 14.359 years that were taken with the same array and at the same frequency, we detected shifts up to 6~mas in the position of the lensed images. We test the possibility that the cause of these shifts is due to a motion of the lensing galaxies, which we find is unlikely. Therefore, we conclude that the observed positional shift seen in the lensed images is due to proper motion in the source plane, which we could reconstruct with a precision down to about 1 $\mu$as~yr$^{-1}$. Such an outstanding precision is more than a factor of 10 better than that of the best optical proper motions currently available \citep{Massari2018, Libralato2018}, and demonstrate the power of combining radio interferometry with gravitational lensing techniques. 

To explain the observed proper motion, we investigated two possible scenarios. Assuming that the source consists of a single AGN, a possible explanation for the proper motion is given by the movement of knots moving along the radio jets. However, the de-magnified flux densities of the components are apparently not consistent with knots moving along an approaching and a receding jet. The second and more exotic scenario consists of two radio-loud AGN separated by $\sim$175~pc in projection, both with a core-jet morphology, which form a DAGN system. In this scenario, which is mainly driven by the motion of the flat-spectrum radio components, the two core-jet AGN are moving relative to each other and the jet components are misaligned. If genuine, identifying a DAGN at redshift 3 would have important implications for our understanding of galaxy formation at high redshift, as it would be evidence in favour of the merging scenario for the formation of SMBHs.

The relative position of the candidate DAGN in MG~B2016+112 depends on the lens mass model. Therefore, any error in the lens model translates to an incorrect estimate of the proper motion. Our lens mass model indicates that there is the presence of an astrometric anomaly, even when the companion satellite galaxy is explicitly taken into account. This implies that the mass density distribution is more complex than the model presented here, which includes the main lensing galaxy and its closest satellite galaxy. For example, a Bayesian grid-based analysis (e.g.  \citealt{Dye2005, Koopmans2005, Vegetti2009}) can help in understanding whether this parametric model is overly simplified. Moreover, by performing pixelated potential corrections to the smooth potential associated with the main lensing galaxy, it will be possible to quantify the mass of the satellite galaxy and better constrain the substructure mass density profile, which in our model is fixed \citep{Vegetti2014}. Also, modelling simultaneously the multi-wavelength extended emission from MG~B2016+112 will give many more observational constraints to the mass model \citep{Suyu2012}.

Together with a more sophisticated lens mass modelling, we also need additional observations to further constrain the source scenarios for MG~B2016+112. Further radio observations at higher angular resolution and dynamic range will improve the precision of the image positions, as they would resolve all the subcomponents in the lensed images, and,  therefore, better determine the relative motions of sources 1 and 2.  Moreover, the detection of the counter-images of region C is a direct test to the validity of the lens mass model presented here. Also, optical spectroscopic observations can potentially reveal velocity offsets between the optical emission lines associated with the AGN activity (e.g. Ly$\alpha$) in the different lensed images, which would add to the case for the DAGN scenario. Spectroscopic observations at high spectral resolution would also clearly discern between the presence of two narrow line regions or multiple photo-ionization levels within a single complex narrow line region associated with MG~B2016+112.

\begin{acknowledgements}
 We thank the anonymous referee for the useful suggestions on this manuscript. CS would like to thank Z.~Paragi, R.~Schultz and R.~Morganti for the useful discussions about this work. CS thanks JIVE for their help with the data reduction of the Epoch 1 observations.
CS and JPM acknowledge support from a NWO-CAS grant (project number 629.001.023). DM acknowledges financial support from a Vici grant from NWO. LVEK is supported through an NWO-VICI grant (project number 639.043.308). 
The European VLBI Network is a joint facility of independent European, African, Asian, and North American radio astronomy institutes. Scientific results from data presented in this publication are derived from the following EVN project code: GP0030. The National Radio Astronomy Observatory is a facility of the National Science Foundation operated under cooperative agreement by Associated Universities, Inc.
   
\end{acknowledgements}

\bibliographystyle{aa} 
\bibpunct{(}{)}{;}{a}{}{,} 
\bibliography{references} 

\end{document}